\documentclass[apj]{emulateapj}
\graphicspath{{figures/}}
\DeclareGraphicsExtensions{.jpg,.pdf,.png,.eps,.ps}

\usepackage[table,usenames,dvipsnames]{xcolor}
\usepackage{amsmath}
\usepackage{subfigure}
\usepackage[backref,breaklinks,colorlinks,citecolor=blue]{hyperref}
\usepackage{natbib}
\usepackage{graphicx}
\usepackage{multirow}
\usepackage{soul}

\newcommand{\lknee}{\ensuremath{\ell_{\rm knee}}}
\newcommand{\lcut}{\ensuremath{\ell_{\rm min}}}

\newcommand{\be}{\begin{equation}}
\newcommand{\ee}{\end{equation}}

\newcommand{\sptnew}{SPT-3G }

\newcommand{\advactpol}{Adv.~ACTpol }

\newcommand{\removed}[1]{\textcolor{Red}{}}
\include{number_list}

\def\Melbourne{1}

\begin{document}

\title{Extreme Digitisation For Ground-Based Cosmic Microwave Background Experiments}
\author{L.~Balkenhol\altaffilmark{\Melbourne} and C.~L.~Reichardt\altaffilmark{\Melbourne}}
\altaffiltext{\Melbourne}{School of Physics, University of Melbourne, Parkville, VIC 3010, Australia}
\email{lbalkenhol@student.unimelb.edu.au}

\begin{abstract}

The large size of the time ordered data of cosmic microwave background experiments presents challenges for mission planning and data analysis. These issues are particularly significant for Antarctica- and space-based experiments, which depend on satellite links to transmit data. We explore the viability of reducing the time ordered data to few bit numbers to address these challenges. Unlike lossless compression, few bit digitisation introduces additional noise into the data. We present a set of one, two, and three bit digitisation schemes and measure the increase in noise in the cosmic microwave background temperature and polarisation power spectra. The digitisation noise is independent of angular scale and is well-described as a constant percentage of the original detector noise. Three bit digitisation increases the map noise level by $< 2\%$, while reducing the data volume by a factor of ten relative to 32-bit floats. Extreme digitisation is a promising strategy for upcoming experiments.

\end{abstract}

\section{Introduction}
\label{sec:intro}

Observations of the cosmic microwave background (CMB) have played a key role in cosmology \citep{penzias1965, smoot1992, bennett2013, sptpol2013, planck2018}. Current ground-based CMB experiments, like \sptnew \citep{benson2014} at the South Pole telescope (SPT) and \advactpol \citep{thornton2016}, at the Atacama cosmology telescope (ACT), target science goals such as the discovery of inflationary gravitational waves, measuring the number of relativistic species, the neutrino mass sum, and mapping the large-scale distribution of matter through gravitational lensing and the Sunyaev-Zeldovich (SZ) effects. Through lensing and the SZ effects the CMB probes structure formation, reionisation, and is a powerful test for dark matter and dark energy models \citep{s4sciencebook, 2018corecos, litebird2016, pixie2011}.


Over the last two decades, CMB experiments have gone from single detectors to over $10,000$ detectors. The CMB community has developed a variety of compression techniques and computational approaches to handle the increasing volume of data \citep{tristam2007}. These include the compression of time-ordered data (TOD) into maps \citep{tegmark1997}, bandpower estimation \citep{tegmark1998}, and the pseudo-$C_l$ method \citep{brown2005}.

A potential hurdle for experiments at remote locations are the transmission limitations of satellite links. Space-based experiments have employed a combination of lossless and lossy compression techniques, including reduced bits in the TOD \citep{gaztanaga1998, maris2003}. Antarctica-based experiments that transmit a portion of their results via a satellite link downsample their data to meet telemetry allocations, but have not yet used few bit digitisation of the TOD. As we approach the next generation ground-based experiment, CMB-S4 \citep{s4sciencebook}, and the launch of the new space-based missions COrE+ \citep{core2018}, LiteBIRD \citep{litebird2016}, and PIXIE \citep{pixie2011}, we must consider potential transmission bottlenecks carefully. 

In this work we present the method of extreme digitisation, which reduces a many bit (often 32 or 64 bit) signal to a few bits for ground-based experiments. We apply extreme digitisation to simulated TOD and detail the resulting effects on temperature and polarisation power spectra. We find that an optimal three bit digitisation scheme adds $<2\%$ to the map noise level.

This work is structured as follows. In section \S\ref{sec:dig} we detail the challenges that come with handling large data volumes, introduce the process of extreme digitisation and lay out the framework used to test its performance. We present results for white and $1/f$ detector noise in section \S\ref{sec:results}. We summarise our findings in \S\ref{sec:conclusions}.







\section{Digitisation}
\label{sec:dig}

\subsection{The Challenges Of Large Data Sets}
\label{subsec:problem}


The science goals of upcoming CMB experiments depend on achieving substantially faster mapping speeds. Given CMB detectors are generally photon noise limited, improving the mapping speed means adding more detectors. As a result the number of detectors (and data volume) of ground-based experiments has followed an exponential trend like Moore's law, doubling approximately every 2 years \citep{s4sciencebook, Abazajian2015}.

The South Pole is one of the best sites for CMB observations on Earth \citep{chamberlin2001, spt2004}. CMB-S4 plans to include several telescopes at the South Pole \citep{s4sciencebook, barron2018}, which will generate a data influx of $\sim\mathcal{O}(10)\mathrm{Tb/d}$. Transferring this data volume via satellite would be expensive. For context, the transmission allocation for a current CMB experiment, SPT-3G, is $150\mathrm{Gb/d}$. The transmission bottleneck could be overcome by recovering the full data on hard drives every summer and transmitting a downsampled version of the data. Downsampling eliminates high frequency information, which makes it unsuitable for science on small angular scales, such as SZ galaxy clusters. The potential delay (i.e. only getting the high frequency data once a year) also introduces risks by delaying when potential faults or issues at high frequency are noticed. One can reduce these issues by running substantial portions of the analysis at the South Pole, but this comes with its own costs and challenges.



Beyond transmission challenges, a larger data volume increases the size and cost of disk arrays and makes end-to-end simulations of experiments for the purpose of optimisation on systematics estimation more time consuming. Given the sheer size of upcoming data sets, full end-to-end simulations may prove impractical for CMB-S4 \citep{s4sciencebook}. The exponential growth of CMB data makes timestream level operations, such as noise removal and map making, increasingly time consuming. Few bit digitisation could ameliorate all of these challenges.



The Planck mission has already demonstrated the success of extreme digitisation for space-based CMB experiments \citep{maris2003}. \cite{jenet1998} explored the application of few bit digitisation to radio pulsar timing measurements. Recently Clearwater et al. (private communication) have demonstrated the advantages of using one and two bit data when searching for continuous gravitational waves using the Laser Interferometer Gravitational-Wave Observatory (LIGO).

\subsection{Extreme Digitisation}
\label{subsec:extremedigitisation}

The optimal digitisation scheme to minimise distortion for a fixed number of bits depends on the details of the input signal. 
In most cases this optimisation is neglected, since the distortions become vanishingly small as the number of bits increases. However, optimal schemes are critical to the success of extreme (few bit) digitisation. We review the key aspects of designing digitisation schemes as established by \cite{max1960} below.

Digitisation discretises an input signal by sorting it into $N$ ranges, such that an input between $x_i$ and $x_{i+1}$ produces an output value $y_i$. The set of parameters $N, x_i, y_i$ fully specify a digitisation scheme. Conventionally one chooses $x_{1} = -\infty$ and $x_{N+1} = \infty$, i.e. values beyond some threshold saturate and yield identical output. In order to quantify the performance of a given digitisation scheme we define the distortion as
\begin{equation}\label{eq:distdef}
D = \left\langle  \left( s - \hat{s} \right)^2 \right\rangle,
\end{equation}
where $s$ is the input and $\hat{s}$ the output signal. For a stochastic input signal we can calculate an amplitude probability distribution $p(x)$. This allows us to re-express the distortion as a sum over the digitisation levels:

\begin{equation} \label{eq:dist}
D = \sum_{i = 1}^N \int_{x_i}^{x_{i+1}} \left(x-y_i\right)^2 p(x) dx.
\end{equation}

Since we wish to minimise the distortion, we set its derivatives with respect to $x_i$ and $y_i$ to zero. Starting with $x_i$ we obtain

\begin{equation} \label{eq:distderiv1}
\frac{\partial D}{\partial x_i} = \left(x_i-y_{i-1}\right)^2 p(x_i) - \left(x_i - y_i\right)^2 p(x_i) = 0.
\end{equation}

\begin{equation} \label{eq:digitequalspacecondition}
x_i = \frac{y_i+y_{i+1}}{2},
\end{equation}
which informs us that the threshold levels should lie midway between their adjacent output levels. Setting the derivative of equation \ref{eq:dist} with respect to $y_i$  to zero gives the additional condition
\begin{equation} \label{eq:distderiv2}
\frac{\partial D}{\partial y_i} = -2 \int_{x_i}^{x_{i+1}} \left( x-y_i \right) p(x) dx = 0.
\end{equation}

\begin{equation} \label{eq:digitareacondition}
\int_{x_i}^{x_{i+1}} \left( x-y_i \right) p(x) dx = 0.
\end{equation}
This condition implies that we should choose $y_i$, such that it halves the area underneath $p(x)$ in the interval from $x_i$ to $x_{i+1}$.

To progress further, we must consider the probability distribution $p(x)$ of the input signal. 
In this work, we are interested in the compression of the TOD from each detector of a ground-based CMB experiment. 
Thus we can safely assume that the input signal is in the low signal-to-noise regime. 
Furthermore, we take the simplest noise profile: Gaussian, white detector noise\footnote{As discussed in the appendix, this treatment also applies to more realistic noise profiles, because the large number of points lying in each map pixel allows the application of the central limit theorem.}. 
Together these assumptions imply that the input, $x$, is normally distributed, with $p(x) = (1/\sqrt{2\pi\sigma^2}) e^{-x^2/2\sigma^2}$. 

\citet{max1960} solved Equations \ref{eq:digitequalspacecondition} and \ref{eq:digitareacondition} for 1-, 2- and 3-bit digitisation for this choice of probability distribution, $p(x)$. 
We quote their results below and refer the reader to the aforementioned work for the derivation.  
As one would expect given the symmetry in $p(x)$, the 1-bit (N=2 output levels) case is simply:
\begin{equation} \label{eq:1bit}
\hat{s}_1(t) = \left\{ \begin{array}{lr}
1, & s(t) > 0\\
-1, & s(t) \leq 0
\end{array} \right. \hspace{0.5cm}.  \end{equation}
Essentially, the optimal 1-bit digitisation scheme applies the sign function to the TOD. 

For two bit digitisation, we can encode $N=4$ output levels. The optimal set of thresholds and digitisation levels is given by the four-level function:
\begin{equation}  \label{eq:2bit}
\hat{s}_2(t) = \left\{ \begin{array}{rl}
1.51 \sigma, & s(t) \geq 0.98 \sigma\\
0.45 \sigma, & 0 \leq s(t) < 0.98 \sigma\\
-0.45 \sigma, & -0.98 \sigma \leq s(t) < 0\\
-1.51 \sigma, & -0.98 \sigma < s(t)\\
\end{array} \right. \hspace{0.5cm}.  \end{equation}

Finally, with 3 bits we may specify $N=8$ output levels. The optimal three bit digitisation is described by the eight-level function:
\begin{equation}  \label{eq:3bit}
\hat{s}_3(t) = \left\{ \begin{array}{rl}
2.15 \sigma, & s(t) \geq 1.75 \sigma\\
1.34 \sigma, & 1.05 \sigma \leq s(t) < 1.75 \sigma\\
0.76 \sigma, & 0.50 \sigma \leq s(t) < 1.05 \sigma\\
0.25 \sigma, & 0 \leq s(t) < 0.50 \sigma\\
-0.25 \sigma, & -0.50 \sigma \leq s(t) < 0\\
-0.76 \sigma, & -1.05 \sigma \leq s(t) < -0.50 \sigma\\
-1.34 \sigma, & -1.75 \sigma \leq s(t) < -1.05 \sigma\\
-2.15 \sigma, & -1.75 \sigma < s(t)\\
\end{array} \right. \hspace{0.5cm}.  \end{equation}

The digitisation schemes above are designed to minimise the distortion of the TOD as defined in Equation \ref{eq:distdef}. This does not ensure that the digitised output has the same signal power as the input. However, the output of the schemes given above can be rescaled by a constant to achieve power conservation. This is demonstrated in the appendix, where we also derive an analytical expression for the normalisation constant. We use this analytical rescaling for all results in this work.




\subsection{From Time Ordered Data To Maps}
\label{subsec:method}

To investigate the performance of the derived digitisation schemes, we simulate timestream level scans over a single CMB realisation. We add detector noise and apply the digitisation schemes above to the simulated I, Q, and U TOD. We also retain the original 64 bit TOD to construct control maps. The different timestreams are binned into HEALPix \citep{healpix}\footnote{Available at \url{http://healpix.sourceforge.net/}.} maps with resolution $\mathrm{NSIDE} = 4096$ for a T, E, and B power spectrum analysis.


We use the healpy Python package to generate a single CMB realisation for the best-fit Planck 2015 cosmology. Specifically, the key cosmological parameters are $H_0 = 67.8 \mathrm{km \> s^{-1} Mpc^{-1}}$, $\Omega_{\mathrm{m}} = 0.308 $, $n_{\mathrm{s}} = 0.968$, and $\tau = 0.066$ \citep{planck2016}.


We simulate observing a $600 \> \mathrm{ deg^2}$ patch of the sky with a single detector. The detector noise level for temperature is $500 \mathrm{\mu K \sqrt{s}}$ and for polarisation observations $\sqrt{2} \hspace{0.7ex} 500 \mathrm{\mu K \sqrt{s}}$, i.e. the corresponding photon noise limit. Constant elevation scans (CES) are performed beginning at right ascension (RA) and declination (DEC) $(0^\circ, 0^\circ)$. After covering $24^\circ30'$ along RA the detector is reset to a RA of $0^\circ$ and an offset in DEC corresponding to the pixel size. Through repetition of CES with constant steps in DEC the survey patch is covered. This imitates the scan strategy of the SPT \citep{schaffer2011}. The entire scan strategy is repeated $100$ times with offsets in the starting RA and DEC up to the size of one pixel, ensuring that pixels are sampled uniformly.

We create naive I,Q, and U maps from the simulated I, Q, and U TOD, i.e. the value of a map pixel is the average of all TOD samples lying within that pixel. We create four maps: one using the original 64 bit TOD and one for each of the digitisation schemes. We save the ouput maps at $800$, $8,000$, $80,000$, $1,024,000$, $10,240,000$ and $102,400,000$ hits per pixel.

\subsection{Power Spectrum Estimation}
\label{subsec:psestimation}

To minimise boundary effects in the subsequent analysis we apodise the observed patch using a cosine mask. For the case of white detector noise we use PolSpice \citep{polspice} to compute $\mathrm{TT}$, $\mathrm{EE}$ and $\mathrm{BB}$ power spectra from output maps. 

For the later considered case of anisotropic noise, we use healpy to calculate the spherical harmonic coefficients $a\ell m$. Since the scan strategy of the simulated observations is similar to the SPT's, smaller values of $m$ correspond to larger angular features along the scan direction \citep{chown2018}. Because of the anisotropic nature of the noise, equally weighting all modes leads to inaccurate power spectra. As a first order approximation to optimal weighting we discount the lowest modes in scan direction.

We rebin the calculated power spectra as mentioned when applicable and assign Fisher matrix error bars \citep{tegmarkdesignersguide1997}.

\begin{figure*}[h!t]\centering
\vspace{-1.5cm}
\includegraphics[width=0.8\textwidth,clip]{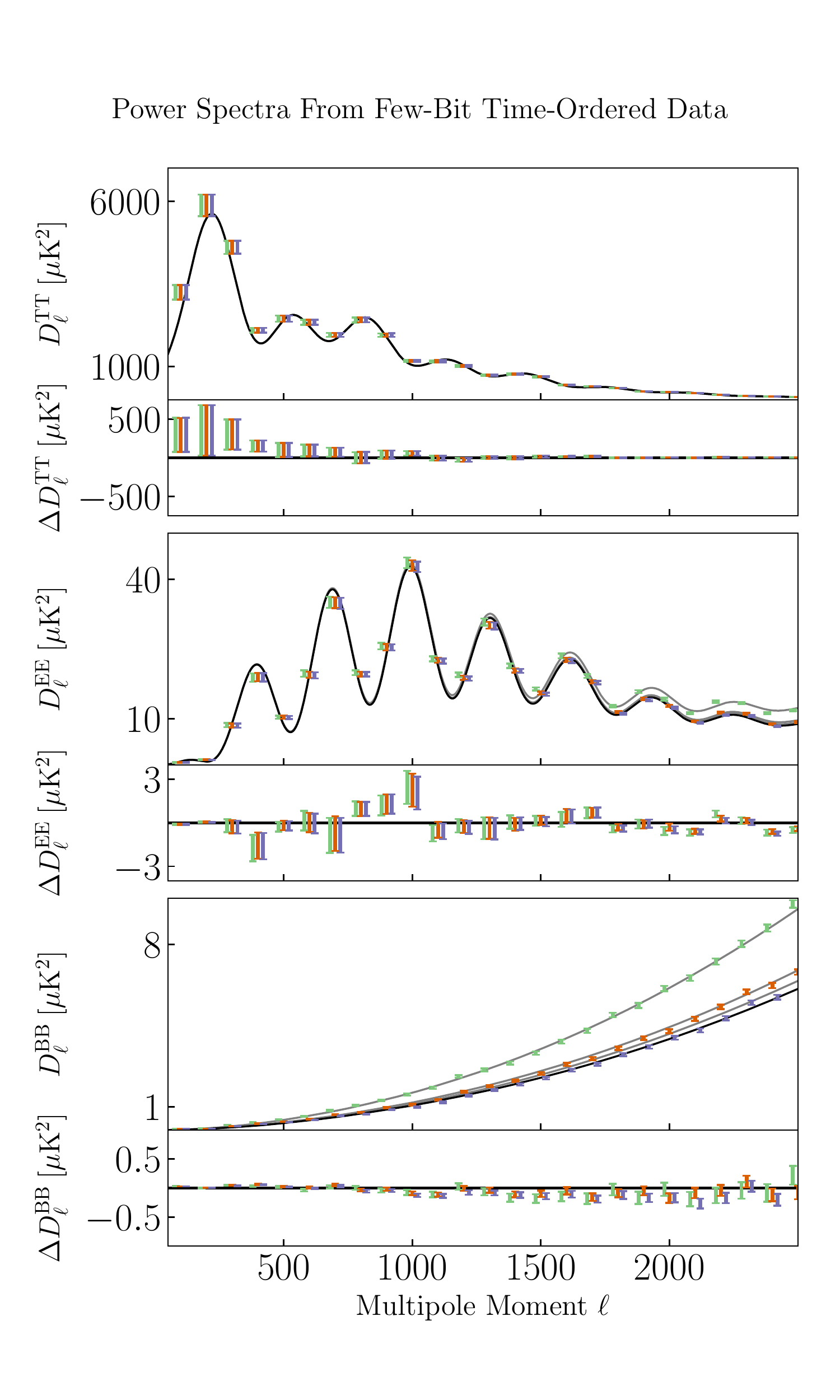}
  \caption[Current ]{
  Input power spectra used to generate the CMB template map plus detector noise level (black line) and digitisation noise (grey line); power spectra using three- (violet marker, right), two- (orange marker, middle), and one-bit (green marker, left) TOD. Power spectra are rescaled to $D_\ell = \ell (\ell+1) C_\ell /2\pi$ and bined to $\Delta \ell = 100$. Main panels show the TT (top), EE (middle), and BB (bottom) power spectra; accompanying panels show the difference between the input plus noise curves and the power spectra recovered from few-bit TOD. Spectra constructed from few-bit TOD recover the input well. This plot shows results for a simulation with $1,024,000$ hits per map pixel using Gaussian white detector noise.}
\label{fig:psrecover}

\end{figure*}

As seen in Figure \ref{fig:psrecover} (appendix) all spectra of extremely digitised data recover the input to a satisfying degree. Differences between few bit and the input power spectra are due to observing a single CMB realisation and residual boundary effects. Non-linearity introduced by the digitisation process disappears in the final map due to the large number of TOD points that contribute to each map pixel. Thus the digitisation effects are not an issue for power spectrum estimation.

\section{Results}
\label{sec:results}

\subsection{White Noise}
\label{subsec:whitenoise}

We analyse the results of the simulated observations to quantify the distortion digitisation causes in the power spectra and investigate whether the additional noise has any angular scale dependence.

We infer the fractional increase of the original map noise level, $\Delta \sigma / \sigma$, to quantify the distortion caused by extreme digitisation. We calculate $\Delta \sigma / \sigma$ from the noise dominated angular scales in the power spectra via

\begin{equation}
\frac{\Delta \sigma}{\sigma} = \sqrt{1 + \frac{C_\ell^\mathrm{X}}{C_\ell^{\mathrm{N}}}} - 1,
\end{equation}

where $C_\ell^{\mathrm{N}}$ and $C_\ell^\mathrm{X}$ are the detector noise level and the additional digitisation noise respectively. Before calculating $C_\ell^\mathrm{X}/C_\ell^\mathrm{N}$ we rebin the power spectra to $\Delta \ell \approx 100$. This ensures that points in the noise tail are independent, allowing us to extract an uncertainty for $\Delta \sigma / \sigma$.







The deduced additional noise for one, two, and three bit digitisation is shown in Table \ref{tab:extranoisewhite} in the appendix. We see that 3 bit digitisation performs best, followed by two bit and one bit. This is expected, since more bits allow a more faithful representation of the input signal. We observe that the fractional increase, $\Delta \sigma / \sigma$, is independent of the hits per pixel in the maps and that the compression performs as well for polarisation as is does for temperature data. Finally, it is striking how little noise is added. On average one-bit digitisation leads to a $(25.13\pm 0.94)\%$ increase in the map noise level, two-bit yields a $(6.40\pm0.41)\%$ increase and three-bit adds only $(1.76\pm0.22)\%$ ($1\sigma$ confidence intervals given. This is impressive, given that all schemes considered reduce the TOD volume by at least an order of magnitude.

\begin{figure}[htb]\centering
\includegraphics[width=0.5\textwidth,clip]{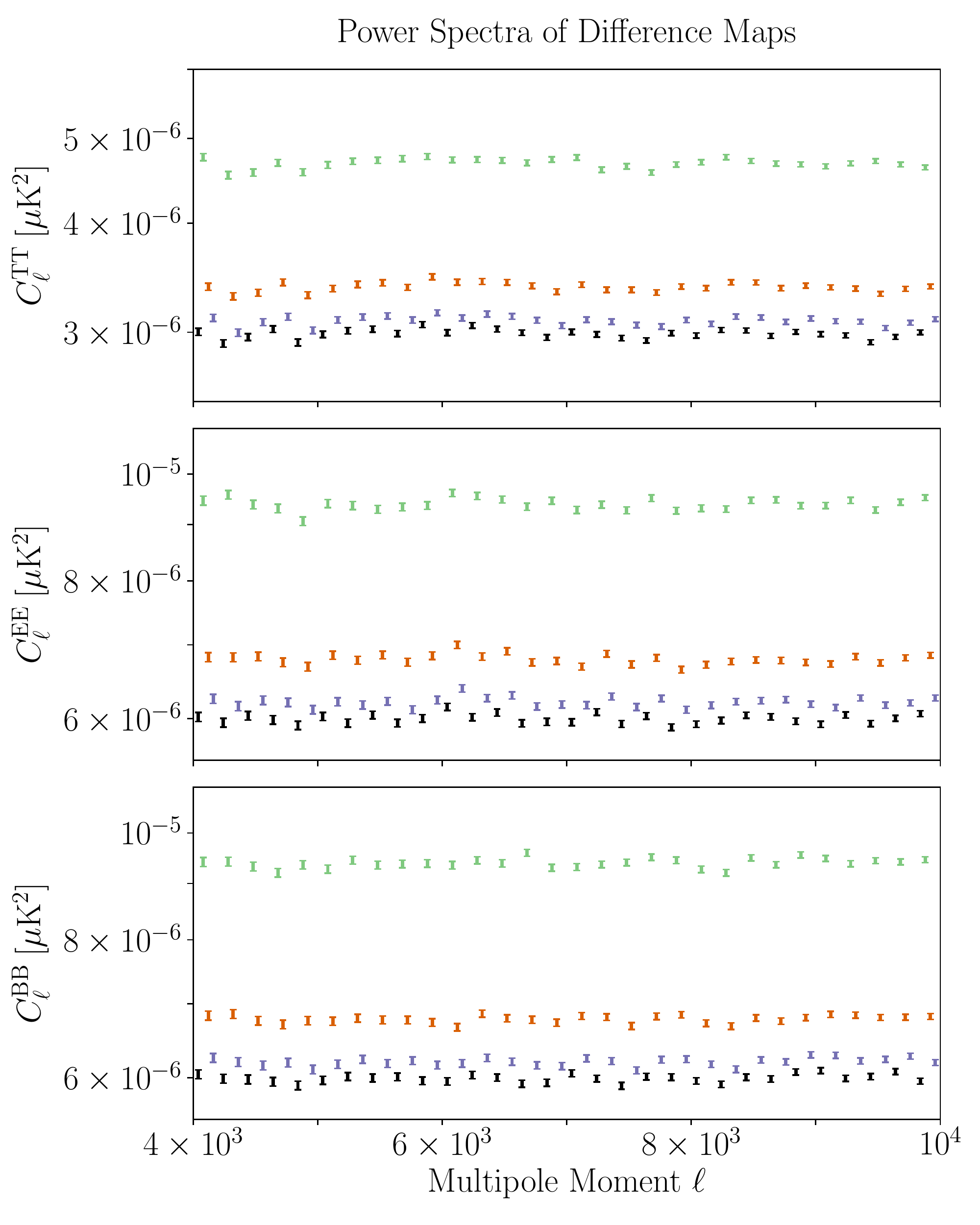}
  \caption[Current ]{
  TT, EE, and BB power spectra (upper, middle and lowest panel respectively) of difference maps for one-bit (green marker, high), two-bit (orange marker, medium), and three-bit digitisation (violet marker, low). The control case using 64-bit TOD is also shown (black marker). All power spectra are rebined to $\Delta \ell = 200$. The vertical axes are logarithmic. The noise increase due to digitisation appears to be independent of angular scale. This plot shows results for a simulation with $1,024,000$ hits per map pixel using Gaussian white detector noise.
\label{fig:diffpswn}
}
\end{figure}

To investigate whether the noise added through digitisation has any angular scale dependence, we subtract the input map from simulated observations. The power in the difference map is plotted in Figure \ref{fig:diffpswn}. There is no sign of any angular scale dependence.


\subsection{1/f Noise}
\label{subsec:oofnoise}

While digitisation clearly works extremely well for white noise, real ground-based CMB experiments face several sources of low-frequency noise. High on the list is atmospheric noise. Modelling the atmosphere as a thin sheet dominated by Kolmogorov turbulence, we can describe fluctuations with the noise profile \citep{lay2000}:

\begin{equation} \label{eq:1/fnoiseform}
\left| N_{1/f}(\ell_{\mathrm{scan}}) \right|^2 = \left| N_{\mathrm{white}} \right|^2 \left[ 1 + \left( \frac{\ell_{\mathrm{scan}}}{\ell_0} \right)^{-3/2} \right]^2,
\end{equation}


where $N_{\mathrm{white}}$ denotes white detector noise. In this expression $\ell_{\mathrm{scan}}$ describes the multipole moment of angular features of corresponding size along the scan direction. The scale at which such modes are amplified appreciably by the second term in equation \ref{eq:1/fnoiseform} is set by $\ell_0$. The exponent of $\ell_{\mathrm{scan}}/\ell_0$ is chosen to model observation at a constant, moderate elevation \citep{lay1997}. In line with the observed noise profile of the SPT we choose $\ell_0 = 1000$ for temperature and $\ell_0 = 0$ for polarisation; this means that power is doubled at $\lknee \approx 1800$ and $\lknee \approx 90$ for temperature and polarisation, respectively \citep{henning2018}. In practice CMB experiments suppress power in the lowest modes through a range of techniques, including fitting polynomials, sines, and cosines to the TOD. We simulate successful removal of these modes by applying a high pass filter to the noise, such that

\begin{equation} \label{eq:1/fnoiseformfinal}
\left| N (\ell_{\mathrm{scan}}) \right|^2 = \left| N_{1/f}(\ell_{\mathrm{scan}}) \right|^2 \exp\left[ -2\left( \frac{\lcut}{\ell_{\mathrm{scan}}} \right)^{6} \right],
\end{equation}

with $\lcut = 336.3$ for temperature and $\lcut = 16.8$ for polarisation.

We use the same digitisation schemes as for white noise, but replace the parameter $\sigma$ in equations \ref{eq:1bit}, \ref{eq:2bit} and \ref{eq:3bit} by considering Parseval's theorem on the above noise profile. The appropriate value is obtained by numerically evaluating the integral

\begin{equation} \label{eq:psvl1/f}
\sigma^2 = \int_0^\infty \left| N(\ell) \right|^2 d\ell.
\end{equation}

After modifying the digitisation schemes derived in subsection \S\ref{subsec:extremedigitisation} accordingly, we carry out the same simulations as described in subsection \S\ref{subsec:method} with the noise profile given in equation \ref{eq:1/fnoiseformfinal}. We save the output maps at $8,000$, $102,400$ and $1,024,000$ hits per pixel. The analysis procedure of subsection \S\ref{subsec:whitenoise} is used to characterise the digitisation noise.


We recover similar results to the white noise case, with three bit digitisation performing the best, followed by two bit, and one bit. Table \ref{tab:extranoiseoof} in the appendix displays the fractional increase to the detector noise level due to digitisation, $\Delta \sigma / \sigma$. While polarisation observations seem to incur a slightly higher noise penalty compared to temperature, deviations do not exceed $3\sigma$. On average one-bit digitisation leads to a $(24.4\pm 1.4)\%$ increase in the map noise level, two-bit adds $(6.21\pm0.50)\%$, and three-bit yields an increase of only $(1.71\pm0.21)\%$ ($1\sigma$ confidence intervals given). Thus the effectiveness of few bit digitisation is not limited to the white noise case, but extends to more realistic noise profiles.



\subsection{Existing Compression Techniques}

Current-generation ground-based CMB experiments employ a combination of lossy, e.g. downsampling, and lossless compression, e.g. FLAC, bzip, LZMA, to manage their transmission bottlenecks. We test whether the data volume of timestreams that have undergone few-bit digitisation can be reduced further by applying the most popular lossless compression algorithm: FLAC. 

We obtain 2 minutes of SPT-3G TOD and apply the digitisation schemes in equations \ref{eq:1bit}-\ref{eq:3bit} before compressing the data further using FLAC. We compare the file size of FLAC compressed few-bit timestreams to the typical data reduction rate achieved for 24-bit SPT-3G data.

We find that using FLAC in conjunction with extreme digitisation does not yield additional data volume reduction. For 3-bit digitisation, both the FLAC and raw timestream have $\sim$3 bits per data point. In fact, due to the overhead required by FLAC compressed files perform marginally worse than uncompressed few-bit streams. This is unsurprising, given that we have already maximally compressed the data using extreme digitisation. For comparison, FLAC reduces the full 24-bit  SPT-3G data to $6-8$ bits of entropy per number. For this real-world case, three-bit digitisation reduces the data volume by a factor of $2-3$ more than lossless compression.

\section{Conclusion}
\label{sec:conclusions}

In this work we have conducted an investigation of extreme digitisation as a technique in combating the challenges of large data volumes for ground-based CMB experiments. In particular the reduction of TOD volume by an order of magnitude decreases the transmission requirements from remote locations. The reduction may also streamline the analysis and simulation of TOD.

We present a set of one, two, and three bit digitisation schemes. For white and $1/f$ detector noise alike, we find that optimal three bit digitisation adds as little as $<2\%$ to the map noise level. This is true for temperature and polarisation observations. No change in the results is observed for maps with different numbers of hits per pixel and no angular scale dependence was observed in the added noise. Applying lossless compression techniques like FLAC to few-bit digitised data does not yield further reductions in data volume.

Given that the digitisation noise remains low at small angular scales, cluster finding algorithms may perform well on few bit TOD. Investigating this will also help determine the higher order statistical moments, i.e. skewness and kurtosis, of the digitisation noise.

\acknowledgments 


We are indebted to Nathan Whitehorn for the details about SPT data compression provided by him.
We are grateful for insightful discussions about the prospects of extreme digitisation for CMB experiments with Andrew Melatos and Patrick Clearwater.
We thank the referee as well as Nikhel Gupta, Srinivasan Raghunathan, Federico Bianchini, Andrew Melatos, and Patrick Clearwater for valuable feedback on the manuscript.
We acknowledge support from an Australian Research Council Future Fellowship (FT150100074), and also from the University of Melbourne. 
This research used resources of the National Energy Research Scientific Computing Center, which is supported by the Office of Science of the U.S. Department of Energy under Contract No. DE-AC02-05CH11231. 
We acknowledge the use of the Legacy Archive for Microwave Background Data Analysis (LAMBDA). Support for LAMBDA is provided by the NASA Office of Space Science.


This research made use of the NumPy \citep{numpy}, SciPy \citep{scipy}, Matplotlib \citep{matplotlib}, and Astropy \citep{astropy} packages.

\newpage

\appendix

\subsection{Digitisation Noise Levels}
\label{subsec:appendixnoisetables}

\def\arraystretch{1.3}
\begin{table*}[tbh]
\begin{center}
\caption{\label{tab:extranoisewhite} Fractional Noise Increase Due To Digitisation - Gaussian White Detector Noise}
\small
\begin{tabular}{c c c c c}
Hits Per Pixel & Channel & 1 Bit & 2 Bit & 3 Bit \\
\hline
\hline
\multirow{3}{*}{800}  & TT  & $ 0.252 \pm 0.0091 $  & $ 0.0639 \pm 0.0034 $  & $ 0.0173 \pm 0.0021 $ \\
& EE  & $ 0.253 \pm 0.012 $  & $ 0.0640 \pm 0.0047 $  & $ 0.0175 \pm 0.0041 $ \\
& BB  & $ 0.254 \pm 0.011 $  & $ 0.0652 \pm 0.0054 $  & $ 0.0180 \pm 0.0039 $ \\
\hline
\multirow{3}{*}{8,000}  & TT  & $ 0.2521 \pm 0.0076 $  & $ 0.0641 \pm 0.0034 $  & $ 0.0174 \pm 0.0017 $ \\
& EE  & $ 0.2519 \pm 0.0074 $  & $ 0.0636 \pm 0.0040 $  & $ 0.0173 \pm 0.0022 $ \\
& BB  & $ 0.252 \pm 0.018 $  & $ 0.0634 \pm 0.0089 $  & $ 0.0176 \pm 0.0036 $ \\
\hline
\multirow{3}{*}{80,000}  & TT  & $ 0.2526 \pm 0.0080 $  & $ 0.0645 \pm 0.0040 $  & $ 0.0181 \pm 0.0016 $ \\
& EE  & $ 0.2506 \pm 0.0093 $  & $ 0.0641 \pm 0.0042 $  & $ 0.0177 \pm 0.0019 $ \\
& BB  & $ 0.250 \pm 0.012 $  & $ 0.0636 \pm 0.0065 $  & $ 0.0179 \pm 0.0032 $ \\
\hline
\multirow{3}{*}{1,024,000}  & TT  & $ 0.2501 \pm 0.0090 $  & $ 0.0633 \pm 0.0031 $  & $ 0.0173 \pm 0.0016 $ \\
& EE  & $ 0.2518 \pm 0.0080 $  & $ 0.0646 \pm 0.0033 $  & $ 0.0179 \pm 0.0018 $ \\
& BB  & $ 0.2520 \pm 0.0082 $  & $ 0.0639 \pm 0.0037 $  & $ 0.0173 \pm 0.0021 $ \\
\hline
\multirow{3}{*}{10,240,000}  & TT  & $ 0.2480 \pm 0.0078 $  & $ 0.0630 \pm 0.0038 $  & $ 0.0175 \pm 0.0018 $ \\
& EE  & $ 0.2526 \pm 0.0077 $  & $ 0.0647 \pm 0.0033 $  & $ 0.0183 \pm 0.0018 $ \\
& BB  & $ 0.252 \pm 0.010 $  & $ 0.0642 \pm 0.0042 $  & $ 0.0177 \pm 0.0020 $ \\
\hline
\multirow{3}{*}{102,400,000}  & TT  & $ 0.2465 \pm 0.0049 $  & $ 0.0629 \pm 0.0019 $  & $ 0.0170 \pm 0.0010 $ \\
& EE  & $ 0.2511 \pm 0.0078 $  & $ 0.0644 \pm 0.0035 $  & $ 0.0175 \pm 0.0018 $ \\
& BB  & $ 0.2523 \pm 0.0086 $  & $ 0.0640 \pm 0.0031 $  & $ 0.0179 \pm 0.0017 $ \\
\hline
\multirow{3}{*}{Average}  & TT  & $ 0.2502 \pm 0.0017 $  & $ 0.0636 \pm 0.0017 $  & $ 0.0175 \pm 0.0017 $ \\
 & EE  & $ 0.2518 \pm 0.0023 $  & $ 0.0643 \pm 0.0023 $  & $ 0.0177 \pm 0.0023 $ \\
 & BB  & $ 0.2519 \pm 0.0027 $  & $ 0.0640 \pm 0.0027 $  & $ 0.0177 \pm 0.0027 $ \\
\end{tabular}
\tablecomments{ 
Fractional addition to the map noise level, $\Delta \sigma / \sigma$, due to one-, two-, and three-bit digitisation for white detector noise with $1\sigma$ confidence intervals. Note that there is no appreciable variation between temperature and polarisation observations, or between different depths of observation. It is striking that three-bit digitisation only leads to a percent-level increase in the map noise level.}
 \normalsize
\end{center}
\end{table*}

\def\arraystretch{1.3}
\begin{table*}[tbh]
\begin{center}
\caption{\label{tab:extranoiseoof} Fractional Noise Increase Due To Digitisation - $1/f$ Detector Noise}
\small
\begin{tabular}{c c c c c}
Hits Per Pixel & Channel & 1 Bit & 2 Bit & 3 Bit \\
\hline
\hline
\multirow{3}{*}{8,000}  & TT  & $ 0.231 \pm 0.017 $  & $ 0.0584 \pm 0.0052 $  & $ 0.0159 \pm 0.0022 $ \\
& EE  & $ 0.250 \pm 0.012 $  & $ 0.0641 \pm 0.0043 $  & $ 0.0177 \pm 0.0019 $ \\
& BB  & $ 0.245 \pm 0.021 $  & $ 0.0623 \pm 0.0067 $  & $ 0.0169 \pm 0.0032 $ \\
\hline
\multirow{3}{*}{102,400}  & TT  & $ 0.235 \pm 0.014 $  & $ 0.0597 \pm 0.0042 $  & $ 0.0166 \pm 0.0019 $ \\
& EE  & $ 0.2503 \pm 0.0091 $  & $ 0.0636 \pm 0.0042 $  & $ 0.0174 \pm 0.0016 $ \\
& BB  & $ 0.249 \pm 0.024 $  & $ 0.0641 \pm 0.0069 $  & $ 0.0178 \pm 0.0025 $ \\
\hline
\multirow{3}{*}{1,024,000}  & TT  & $ 0.238 \pm 0.013 $  & $ 0.0601 \pm 0.0046 $  & $ 0.0168 \pm 0.0019 $ \\
& EE  & $ 0.2502 \pm 0.0078 $  & $ 0.0630 \pm 0.0039 $  & $ 0.0173 \pm 0.0019 $ \\
& BB  & $ 0.2499 \pm 0.0095 $  & $ 0.0636 \pm 0.0049 $  & $ 0.0177 \pm 0.0019 $ \\
\hline
\multirow{3}{*}{Average}  & TT  & $ 0.2346 \pm 0.0020 $  & $ 0.0594 \pm 0.0020 $  & $ 0.0164 \pm 0.0020 $ \\
 & EE  & $ 0.2503 \pm 0.0018 $  & $ 0.0636 \pm 0.0018 $  & $ 0.0175 \pm 0.0018 $ \\
 & BB  & $ 0.2480 \pm 0.0025 $  & $ 0.0633 \pm 0.0025 $  & $ 0.0175 \pm 0.0025 $ \\
\end{tabular}
\tablecomments{ 
Fractional addition to the map noise level, $\Delta \sigma / \sigma$, due to one-, two-, and three-bit digitisation for $1/f$ detector noise with $1\sigma$ confidence intervals. As in the white noise case, we observe no significant trend between temperature and polarisation or for different depths of observation. We note that, even assuming a more realistic noise profile, three-bit digitisation only leads to a percent-level increase in the map noise level.}
 \normalsize
\end{center}
\end{table*}

\subsection{Preserving Power in Digitised Data}
\label{subsec:appendixpreservepower}

As mentioned in subsection \S\ref{subsec:extremedigitisation}, equation \ref{eq:distdef} does not demand that a digitisation scheme conserves power. However, it is relatively simple to impose power conservation, since it boils down to rescaling the digitised output by a constant, $\gamma$. We derive an expression for $\gamma$, such that a cross-spectrum of two digitised maps yields an unbiased estimate of the input power spectrum. Mathematically, this is equivalent to:

\begin{equation} \label{eq:normcrosspower}
\langle (\mu + \xi_1) (\mu + \xi_2) \rangle = \gamma^2 \langle D(\mu + \xi_1) D(\mu + \xi_2) \rangle,
\end{equation}

where all timestreams sample the same underlying signal, $\mu$, but add different noise realisations $\xi_{1, 2}$. The digitisation process is denoted by $D(\dots)$. The covariance of two timestreams must be zero. This remains true for the digitised signals, as digitisation discretises the probability distributions at play but does not change the underlying dynamics. Therefore

\begin{equation}
\langle D(\mu + \xi_1) D(\mu + \xi_2) \rangle = \sum_1 \sum_2 D(\mu + \xi_1) D(\mu + \xi_1) P(D(\mu + \xi_1)) P(D(\mu + \xi_2)),
\end{equation}

where we sum over all output levels of both timestreams $(1, 2)$ and $P$ denotes the probability of an output level occurring. For current-generation CMB experiments it is common to draw from $>10^5$ samples to obtain an estimate of a map pixel. This corresponds to adding many noise realisation on top of the same signal, $\mu$. 
The central limit theorem ensures that the amplitude probability distribution function of the noise in each pixel is Gaussian. 
The probabilities $P$ in the above expression can therefore be expressed as integrals over normal distributions, $p(\xi)$, with mean zero and standard deviation $\sigma$.

\begin{equation}
\langle D(\mu + \xi_1) D(\mu + \xi_2) \rangle = \sum_{i=0}^N \sum_{j=0}^N  \int_{x_i-\mu}^{x_{i+1}-\mu} \int_{x_j-\mu}^{x_{j+1}-\mu} y_i y_j p(\xi_1) p(\xi_2) d\xi_1 d\xi_2.
\end{equation}

We separate the above sums and recognise the error function.

\begin{equation}
\begin{aligned}
\langle D(\mu + \xi_1) D(\mu + \xi_2) \rangle &= \left\{ \sum_{i=0}^N  \frac{y_i}{2} \left[ erf \left( \frac{x_{i+1} - \mu}{\sqrt{2}\sigma} \right) - erf \left( \frac{x_{i} - \mu}{\sqrt{2}\sigma} \right) \right] \right\} \\
&\times \left\{  \sum_{j=0}^N \frac{y_j}{2} \left[ erf \left( \frac{x_{j+1} - \mu}{\sqrt{2}\sigma} \right) - erf \left( \frac{x_{j} - \mu}{\sqrt{2}\sigma} \right) \right] \right\}.
\end{aligned}
\end{equation}

\begin{equation}
\begin{aligned}
&= \left\{ \sum_{i=0}^N  \frac{y_i}{2} \left[ \frac{2}{\sqrt{\pi}} \sum_{n = 0}^\infty \frac{(-1)^n}{n! (2n+1)} \frac{1}{(\sqrt{2}\sigma)^{2n+1}} \left( (x_{i+1}-\mu)^{2n+1} - (x_{i}-\mu)^{2n+1} \right) \right] \right\} \\
&\times \left\{ \sum_{j=0}^N  \frac{y_j}{2} \left[ \frac{2}{\sqrt{\pi}} \sum_{m = 0}^\infty \frac{(-1)^m}{m! (2m+1)} \frac{1}{(\sqrt{2}\sigma)^{2m+1}} \left( (x_{j+1}-\mu)^{2m+1} - (x_{j}-\mu)^{2m+1} \right) \right] \right\}.
\end{aligned}
\end{equation}

\begin{equation}
\begin{aligned}
&= \left\{ \sum_{i=0}^N  \frac{y_i}{\sqrt{pi}} \left[ \sum_{n = 0}^\infty \frac{(-1)^n}{n! (2n+1)} \frac{1}{(\sqrt{2}\sigma)^{2n+1}} \left( \sum_{k=0}^{2n+1} {2n+1 \choose k} \mu^k ( x_{i+1}^{2n+1-k} - x_{i}^{2n+1-k} ) \right) \right] \right\} \\
&\times \left\{ \sum_{j=0}^N  \frac{y_j}{\sqrt{2}} \left[ \sum_{m = 0}^\infty \frac{(-1)^m}{m! (2m+1)} \frac{1}{(\sqrt{2}\sigma)^{2m+1}} \left( \sum_{l=0}^{2m+1} {2m+1 \choose l} \mu^l ( x_{j+1}^{2m+1-l} - x_{j}^{2m+1-l} ) \right) \right] \right\} ,
\end{aligned}
\end{equation}

where we have used the Maclaurin series expansion of the error function and the binomial expansion. Moving the $i, j$ summation forwards yields

\begin{equation} \label{eq:dcompleteklsum}
\begin{aligned}
\langle D(\mu + \xi_1) D(\mu + \xi_2) \rangle &=  \frac{1}{\pi} \sum_{n,m = 0}^\infty \frac{1}{n! (2n+1)} \frac{(-1)^{n+m}}{m! (2m+1)} \frac{1}{(\sqrt{2}\sigma)^{2n+2m+2}} \\
&\times \sum_{k, l = 0}^{2n+1} {2n+1 \choose k} {2m+1 \choose l} \mu^{k+l} \sum_{i,j=0}^N y_i y_j \psi^{k, n}_i \psi^{l, m}_j,
\end{aligned}
\end{equation}

with

\begin{equation}
\psi_i^{k,n} = x_{i+1}^{2n+1-k} - x_{i}^{2n+1-k}.
\end{equation}

We split the innermost term in equation \ref{eq:dcompleteklsum} and shift indicies such that

\begin{equation}
\sum_{i,j = 0}^N y_i y_j \psi_i^{k, n} \psi_j^{l, m} = \left( \sum_{i = 0}^{N/2}\sum_{j = 0}^{N/2} + \sum_{i = N/2}^{N}\sum_{j = 0}^{N/2} + \sum_{i = 0}^{N/2}\sum_{j = N/2}^{N} + \sum_{i = N/2}^{N}\sum_{j = N/2}^{N} \right) y_i y_j \psi_i^{k, n} \psi_j^{l, m}.
\end{equation}

\begin{equation}
= \sum_{i = 0}^{N/2}\sum_{j = 0}^{N/2} \left( y_i y_j \psi_i^{k, n} \psi_j^{l, m} +  y_{N-i} y_j \psi_{N-i}^{k, n} \psi_j^{l, m} + y_i y_{N-j} \psi_i^{k, n} \psi_{N-j}^{l, m} + y_{N-i} y_{N-j} \psi_{N-i}^{k, n} \psi_{N-j}^{l, m} \right).
\end{equation}

Equations \ref{eq:digitequalspacecondition} and \ref{eq:digitareacondition} in subsection \S\ref{subsec:extremedigitisation} demand that for a Gaussian input distribution $y_i = -y_{N-i}$ and $x_i = -x_{N+1-i}$. This simply states symmetry of the digitisation thresholds and output levels about zero (or after a global mean has been subtracted). Therefore

\begin{equation} \label{eq:ijsumpsiflip}
\sum_{i,j = 0}^N y_i y_j \psi_i^{k, n} \psi_j^{l, m} = \sum_{i = 0}^{N/2}\sum_{j = 0}^{N/2} y_i y_j \left( \psi_i^{k, n} \psi_j^{l, m} + (-1)^{k+1} \psi_{i}^{k, n} \psi_j^{l, m} + (-1)^{l+1} \psi_i^{k, n} \psi_{j}^{l, m} + (-1)^{k+l} \psi_{N-i}^{k, n} \psi_{N-j}^{l, m} \right),
\end{equation}

because $\psi_{N-i}^{k,n} = (-1)^{k} \psi_{i}^{k,n}$. Since ground-based CMB experiments operate at low signal-to-noise we calculate the leading order terms of the signal, $\mu$, in equation \ref{eq:dcompleteklsum}. We split the sum over $k, l$ in said equation as follows:

\begin{equation}
\sum_{i,j = 0}^N y_i y_j \psi_i^{k, n} \psi_j^{l, m} = \sum_{i = 0}^{N/2}\sum_{j = 0}^{N/2} y_i y_j  \left( \psi_i^{k, n} \psi_j^{l, m} - \psi_{i}^{k, n} \psi_j^{l, m} - \psi_i^{k, n} \psi_{j}^{l, m} + \psi_{i}^{k, n} \psi_{j}^{l, m} \right) = 0 .
\end{equation}

For odd $k$ and $l$ equation \ref{eq:ijsumpsiflip} may be simplified to

\begin{equation}
\sum_{i,j = 0}^N y_i y_j \psi_i^{k, n} \psi_j^{l, m} = \sum_{i = 0}^{N/2}\sum_{j = 0}^{N/2} y_i y_j \left( \psi_i^{k, n} \psi_j^{l, m} + \psi_{i}^{k, n} \psi_j^{l, m} + \psi_i^{k, n} \psi_{j}^{l, m} + \psi_{i}^{k, n} \psi_{j}^{l, m} \right) = 4 \sum_{i = 0}^{N/2}\sum_{j = 0}^{N/2} y_i y_j \psi_i^{k, n} \psi_j^{l, m}.
\end{equation}

This term survives and leads to even powers of $\mu$ in equation \ref{eq:dcompleteklsum}. Considering the case of odd $k$ and even $l$ we observe

\begin{equation}
\sum_{i,j = 0}^N y_i y_j \psi_i^{k, n} \psi_j^{l, m} = \sum_{i = 0}^{N/2}\sum_{j = 0}^{N/2} y_i y_j \left( \psi_i^{k, n} \psi_j^{l, m} + \psi_{i}^{k, n} \psi_j^{l, m} - \psi_i^{k, n} \psi_{j}^{l, m} - \psi_{i}^{k, n} \psi_{j}^{l, m} \right) = 0
\end{equation}

and similarly

\begin{equation}
\sum_{i,j = 0}^N y_i y_j \psi_i^{k, n} \psi_j^{l, m} = \sum_{i = 0}^{N/2}\sum_{j = 0}^{N/2} y_i y_j \left( \psi_i^{k, n} \psi_j^{l, m} - \psi_{i}^{k, n} \psi_j^{l, m} + \psi_i^{k, n} \psi_{j}^{l, m} - \psi_{i}^{k, n} \psi_{j}^{l, m} \right) = 0\\
\end{equation}

for odd $k$ and even $l$. Therefore all odd powers of $\mu$ in equation \ref{eq:dcompleteklsum} vanish. Moreover there is no $\mu$ independent term, since this could only be produced by $k=l=0$, but even $k$ and $l$ terms are shown to vanish. Therefore

\begin{equation}
\langle D(\mu + \xi_1) D(\mu + \xi_2) \rangle = c \mu^2 + \mathcal{O}(\mu^4)
\end{equation}

where $c$ is some constants that depends on the specific digitisation scheme chosen. Returning to equation \ref{eq:normcrosspower} we see

\begin{equation}
\gamma^2  = \frac{\langle (\mu + \xi_1) (\mu + \xi_2) \rangle}{\langle D(\mu + \xi_1) D(\mu + \xi_2) \rangle} = \frac{1}{c + \mathcal{O}(\mu^2)} = \frac{1}{c},
\end{equation}

where we have ignored terms beyond quadratic order. By investigating $k=l=1$ terms in equation \ref{eq:dcompleteklsum} we obtain an expression for $\gamma$.

\begin{equation}
\begin{aligned}
c = \sum_{i,j=0}^N  \frac{y_i y_j}{\pi} \left[ \sum_{n,m = 0}^\infty \frac{1}{n! (2n+1)} \frac{(-1)^{n+m}}{m! (2m+1)} \frac{1}{(\sqrt{2}\sigma)^{2n+2m+2}} \left( {2n+1 \choose 1} {2m+1 \choose 1} ( x_{i+1}^{2n} - x_{i}^{2n} ) ( x_{j+1}^{2m} - x_{j}^{2m} ) \right) \right] .
\end{aligned}
\end{equation}

\begin{equation}
= \sum_{i,j=0}^N \frac{y_i y_j}{2\pi\sigma^2} \left[ \sum_{n = 0}^\infty \frac{(-1)^n}{n!} \frac{1}{(\sqrt{2}\sigma)^{2n}}  ( x_{i+1}^{2n} - x_{i}^{2n} ) \right] \left[ \sum_{m = 0}^\infty \frac{(-1)^m}{m!} \frac{1}{(\sqrt{2}\sigma)^{2m}} ( x_{j+1}^{2m} - x_{j}^{2m} ) \right].
\end{equation}

\begin{equation}
= \sum_{i,j=0}^N \frac{y_i y_j}{2\pi\sigma^2} \left\{ exp \left[ - \left(\frac{x_{i+1}}{\sqrt{2}\sigma} \right)^2 \right] - exp \left[ - \left(\frac{x_{i}}{\sqrt{2}\sigma} \right)^2 \right] \right\} \left\{ exp \left[ - \left(\frac{x_{j+1}}{\sqrt{2}\sigma} \right)^2 \right] - exp \left[ - \left(\frac{x_{j}}{\sqrt{2}\sigma} \right)^2 \right] \right\} .
\end{equation}

The normalisation constant therefore is given by

\begin{equation}
\gamma = \left( \sum_{i,j=0}^N \frac{y_i y_j}{2\pi\sigma^2} \left\{ exp \left[ - \left(\frac{x_{i+1}}{\sqrt{2}\sigma} \right)^2 \right] - exp \left[ - \left(\frac{x_{i}}{\sqrt{2}\sigma} \right)^2 \right] \right\} \left\{ exp \left[ - \left(\frac{x_{j+1}}{\sqrt{2}\sigma} \right)^2 \right] - exp \left[ - \left(\frac{x_{j}}{\sqrt{2}\sigma} \right)^2 \right] \right\} \right)^{-1/2} .
\end{equation}

This can be calculated for a given digitisation scheme, specified by $N, x_i, y_i$, and standard deviation of the noise, $\sigma$. Notice that if the digitisation scheme is chosen such that the digitisation thresholds and output levels depend linearly on the noise level, the normalisation constant is independent of the noise level.


\bibliography{digitisation}

\end{document}